# Effects of hot phonons and thermal stress in micro-Raman spectra of molybdenum disulfide

Peter Sokalski,[1] Zherui Han,[2] Gabriella Coloyan Fleming,[1] Brandon Smith,[1] Sean E. Sullivan,[3] Rui Huang,[4] Xiulin Ruan,[2] and Li Shi[1,3,a)]

## AFFILIATIONS

[1]Walker Department of Mechanical Engineering, The University of Texas at Austin, Austin, Texas 78712, USA
[2]School of Mechanical Engineering and the Birck Nanotechnology Center, Purdue University, West Lafayette, Indiana 47907, USA
[3]Materials Science and Engineering Graduate Program, The University of Texas at Austin, Austin, Texas 78712, USA
[4]Department of Aerospace Engineering and Engineering Mechanics, The University of Texas at Austin, Austin, Texas 78712, USA

Note: This paper is part of the APL Special Collection on Phononics of Graphene, Layered Materials, and Heterostructures.
a)Author to whom correspondence should be addressed: lishi@mail.utexas.edu

## ABSTRACT

Micro-Raman spectroscopy has become an important tool in probing thermophysical properties in functional materials. Localized heating by the focused Raman excitation laser beam can produce both stress and local nonequilibrium phonons in the material. Here, we investigate the effects of hot optical phonons in the Raman spectra of molybdenum disulfide and distinguish them from those caused by thermally induced compressive stress, which causes a Raman frequency blue shift. We use a thermomechanical analysis to correct for this stress effect in the equivalent lattice temperature extracted from the measured Raman peak shift. When the heating Gaussian laser beam is reduced to 0.71 $\mu$m, the corrected peak shift temperature rise is 17% and 8%, respectively, higher than those determined from the measured peak shift and linewidth without the stress correction, and 32% smaller than the optical phonon temperature rise obtained from the anti-Stokes to Stokes intensity ratio. This nonequilibrium between the hot optical phonons and the lattice vanishes as the beam width increases to 1.53 $\mu$m. Much less pronounced than those reported in prior micro-Raman measurements of suspended graphene, this observed hot phonon behavior agrees with a first-principles based multitemperature model of overpopulated zone-center optical phonons compared to other optical phonons in the Brillouin zone and acoustic phonons of this prototypical transition metal dichalcogenide. The findings provide detailed insight into the energy relaxation processes in this emerging electronic and optoelectronic material and clarify an important question in micro-Raman measurements of thermal transport in this and other two-dimensional materials.

Transition metal dichalcogenides (TMDs) provide atomically thin building blocks that are stacked to form emerging heterostructured devices.[1–3] Performance of these electronic and optoelectronic devices is governed by the relaxation rates of the charge and lattice excitations in the two-dimensional (2D) hexagonal layered materials. The energy relaxation processes in TMDs share some similarities with those observed in group IV and III–V semiconductors as well as graphene. Hot electrons and hot optical phonons are present in electronic devices fabricated with the cubic-phase semiconductors[4] because the coupling between the electronic excitation and optical phonon modes is much stronger than those between the acoustic phonon polarizations and either the electronic states or the optical phonons. In suspended graphene, similarly, hot charge carriers thermalize with optical phonons much more quickly than their energy relaxation with acoustic phonon polarizations. In particular, relaxation with the out-of-plane polarized flexural phonon modes is slow due to a restrictive selection rule caused by reflection symmetry and prohibits scattering processes involving an odd number of flexural phonons.[5–7] These different relaxation rates result in underpopulated acoustic phonon populations in graphene devices.[8–10] In comparison, TMDs with broken reflection symmetry are not subjected to this selection rule and additionally exhibit larger exciton binding energies than graphene and cubic-phase semiconductors.[11] In molybdenum disulfide ($MoS_2$), a prototypical TMD, the lifetimes of photoexcited excitons measured by ultrafast pump–probe, ranges from as low as 50 ps in monolayer $MoS_2$ with a direct bandgap[12] to greater than 1 ns in the bulk with an indirect gap.[13] The interactions of the long-lifetime excitons with free carriers in the conduction and valence bands, phonons, and defects may give rise to unique nonequilibrium phenomena in bulk $MoS_2$ when the characteristic time or length scales are reduced to the relevant thermalization time or length scales between different types of energy carriers.

Micro-Raman spectroscopy has been used to probe thermal transport and the nonequilibrium dynamics among the charge carriers and phonons in graphene[10,14,15] and other materials. The Raman peak shift has been commonly used to obtain a peak shift temperature ($T_\omega^{eq}$) that serves as a thermometer of the lattice temperature ($T_L$). In graphene, the frequency of Raman-active phonon modes downshifts with increasing temperature because of thermal expansion and anharmonic decay of the Raman-active modes into intermediate and low-frequency modes.[16] Thus, the Raman peak shift can be used to monitor the equivalent lattice temperature of the Raman-inactive phonon modes that are coupled to the Raman-active mode. Meanwhile, the anti-Stokes (AS) to Stokes intensity ratio is controlled by the Bose–Einstein statistics of the Raman-active mode and can be used to extract the equivalent temperature of the Raman-active optical phonon mode, ($T_{op}$). These two types of Raman thermometers have been used previously to verify the theoretical prediction of underpopulated flexural phonons and other acoustic modes compared to the hot optical modes in suspended graphene irradiated by the submicrometer Raman laser beam.[10]

Attempts have also been made to employ micro-Raman spectroscopy to probe thermal transport and hot phonons in TMDs such as $MoS_2$,[17,18] including relatively thick multi-layer $MoS_2$ samples with interlayer rotational disorders.[19] Compared to suspended monolayer graphene where stress introduced by Raman laser heating can be relaxed, micro-Raman measurements of multi-layer TMDs may result in a compressive thermal stress inside the focused laser spot. The Raman peak frequency is sensitive to stress, so that the Raman spectra have been used as a sensor of mechanically introduced stress when the material is at thermal equilibrium with the environment.[20,21] It remains to be understood how the presence of both nonequilibrium temperature distribution and thermal stress affects the use of the Raman peak position, linewidth, and intensity ratio to study thermal transport and hot phonons in relatively thick multi-layer TMDs.

Here, we report an investigation of the effects of thermal stress and hot phonons on the micro-Raman spectra of suspended $MoS_2$ flakes. We find that the compressive thermal stress in the $MoS_2$ heated by a 0.71 $\mu$m radius Gaussian laser spot reduces the downshift magnitude of the Raman peak by almost 23% compared to the case of uniform sample heating. With this stress effect accounted for, the lattice temperature extracted from the peak shift with a 0.71 $\mu$m Gaussian beam width is just 8% higher than the uncorrected linewidth temperature and 32% lower than the zone-center optical phonon temperature obtained from the intensity ratio. The observed degree of non-equilibrium between the zone-center optical modes and the lattice is much less pronounced than those found previously for suspended monolayer graphene[10] and diminishes when the laser spot size is increased to about 1.5 $\mu$m, in agreement with a multitemperature model.[22]

The $MoS_2$ flake samples were mechanically exfoliated and transferred by a standard polymer-based dry-stamp transfer method[23] over the periodic arrays of 30-$\mu$m-diameter holes on a 200 nm thick silicon nitride (SiN) membrane, which was coated with 5 nm chrome (Cr) and 45 nm heat-spreading gold (Au), as shown in Figs. 1(a) and 1(b). The silicon substrate below the SiN membrane was fixed to a heating stage with a thermocouple adhered to the edge of the silicon substrate and placed in a vacuum chamber that was evacuated with a turbomolecular pump. The Raman laser beam was arranged in a backscattering configuration and focused by either a 50× or 100× achromatic objective lens onto the center of the suspended $MoS_2$. The incident focused laser beam profile and size can be changed by varying the fill ratio of the entrance pupil of the objective.[24] The incident Gaussian beam width $w_o$ was determined by fitting the beam intensity profile measured by a charge-coupled device (CCD) camera with a Gaussian function $\exp(-\frac{2r^2}{w_o^2})$, where $r$ is the distance from the beam center. Within the laser power range reported here, no visual change of the sample surface and no irreversible change of the Raman spectra were observed, indicating a negligible effect of laser-assisted surface modification[25] in the relatively thick sample.

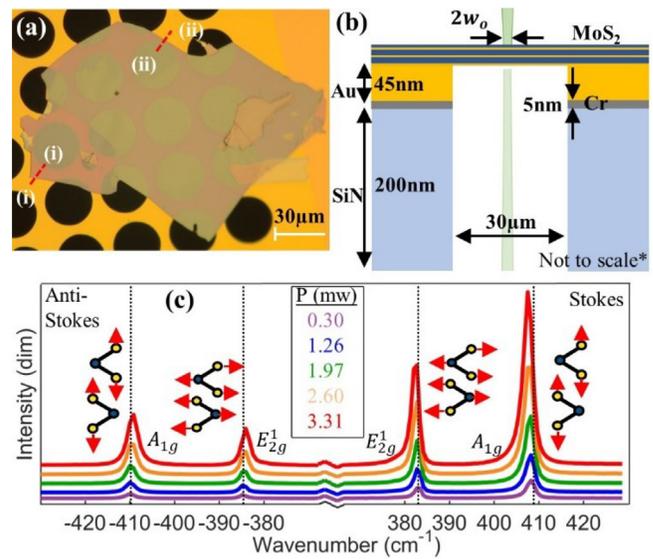

**FIG. 1.** (a) Optical micrograph of a $MoS_2$ flake stamped onto a 200-nm-thick SiN membrane with periodic 30-$\mu$m-diameter holes. The flake thickness is 50 and 72 nm along the $i-i$ and $ii-ii$ hatched lines, respectively. (b) Schematic of the experimental setup for micro-Raman measurements of a suspended $MoS_2$ flake. (c) Micro-Raman spectra of both the Stokes and anti-Stokes $E_{2g}^1$ and $A_{1g}$ bands of $MoS_2$ together with the corresponding atomic displacement vectors. The dashed lines are visual guides demonstrating predominately red-shifting peaks with increasing laser power (P) specified in the legend. The stage temperature is 300 K. A 50× objective was used, resulting in a Gaussian laser beam radius is 1.15 $\mu$m.

Under 532 nm excitation, the Raman spectra [Fig. 1(c)] of the $MoS_2$ flake contain two pronounced $E_{2g}^1$ and $A_{1g}$ optical phonon bands near 383 and 408 cm$^{-1}$, respectively. The Stokes and anti-Stokes $A_{1g}$ bands in $MoS_2$ correspond, respectively, to the absorption and emission of the out-of-plane polarized optical phonons at the zone-center or ZO mode. The measured $E_{2g}^1$ band is attributed to the degenerate in-plane polarized transverse and longitudinal optical (TO and LO) phonons at the zone-center.[26,27] The average between the measured anti-Stokes and Stoke peak positions of each Raman band are used to determine the corresponding peak position and correct for the zero-point offset of the Raman spectrometer. As the laser power increases, the integrated area intensity of the anti-Stokes band increases relative to the Stokes band together with a red shift of the peak frequencies of the two bands and broadening of the Raman linewidth, as shown in Figs. 2(a)–2(c).

The intensity of each of the two Raman bands depends on the Bose–Einstein occupation $n_p(T_{op})$ of the corresponding optical phonon mode $p$ at an equivalent mode temperature $T_{op}$. Because the intensities of the Stokes (S) and anti-Stokes (AS) bands are proportional to $n_p(T_{op}) + 1$ and $n_p(T_{op})$, respectively, the temperature of the optical phonon mode $p$ can be determined from the ratio of the intensities,

$$\frac{I_{AS}}{I_S} = C \frac{(\omega_L + \omega_P)^4}{(\omega_L - \omega_P)^4} \exp\left(\frac{-\hbar\omega_P}{k_b T_{op}}\right), \quad (1)$$

where $\omega_L$ is the laser frequency, $\omega_p$ is the measured zone center optical phonon frequency, and $\hbar$ and $k_b$ are the reduced Planck's constant and the Boltzmann constant, respectively. The factor $C$ depends on the collection efficiencies of each optical setup and the objective lens and must be calibrated for, as discussed in prior works.[8,28] Figure 2(a) shows the laser power-dependent anti-Stokes to Stokes intensity ratios of both the $E_{2g}^1$ and $A_{1g}$ modes at the lowest stage temperature of 300 K. Because the optical phonon temperature equals the known stage temperature at zero laser heating, a second order extrapolation of the data in Fig. 2(a) to zero laser power is used to obtain the intensity ratio at zero laser power and the calibration factor $C$. Unlike graphene,[10] the low optical phonon energy in MoS$_2$ leads to apparent anti-Stokes peaks at room temperature, so that the optical phonon temperatures can be determined for all laser power and stage temperature conditions of this measurement.

As the Raman linewidth is sensitive to the population of phonons that are anharmonically coupled to the Raman-active modes, it can be used to extract a temperature $(T_\Gamma^{eq})$ that may serve as an indictor of the lattice temperature.[29] Meanwhile, the Raman peak shift has been used to probe a local equivalent lattice temperature because it contains aggregated information about the occupation and equivalent temperature of the intermediate frequency phonons (IFPs) and low frequency phonon groups that anharmonically interact with the Raman-active optical phonon modes. This effect explicitly originates from the change of the phonon self-energy arising from anharmonic coupling to growing populations of IFPs and low frequency phonons,[14,16,30] while an implicit contribution due to thermal expansion caused by anharmonicity also manifests as a shift of phonon frequency.[31,32] Using a second-order polynomial fit, the laser power-dependent peak data of the $E_{2g}^1$ and $A_{1g}$ modes in MoS$_2$ are extrapolated to zero laser heating, as shown in Figs. 2(b) and 2(c). The obtained values are the stage temperature-dependent Raman linewidth ($\Gamma_o$) and Raman shift ($\omega_o$) in which all phonons in MoS$_2$ are in equilibrium with the stage temperature and the stress ($\sigma$) vanishes. Variation of the laser spot size is expected to influence the as-obtained $\Gamma_o$ due to a change in the instrument linewidth resolutions, but not $\omega_o$. Figures 2(d)–2(f) show the increasing linewidth and downshifting $\omega_o$ with increasing stage temperature. The latter of which can be used to obtain the temperature coefficient $\left(\frac{\partial \omega}{\partial T}\right)_{\sigma=0}$.

Figure 3 shows the measured $A_{1g}$ and $E_{2g}^1$ peak shifts and linewidths with respect to the optical phonon temperature rise $\Delta T_o$ obtained from the intensity ratio. As the laser spot width decreases from 1.53 to 0.71 $\mu$m, the measured peak red shift and linewidth broadening become apparently smaller than the $\omega_o$ and $\Gamma_o$ that are extracted for an optical phonon temperature to equal the lattice temperature. This behavior can be caused by either hotter optical phonons than the lattice or the presence of a compressive stress ($\sigma$) that is induced by the nonuniform temperature rise ($\Delta T$) due to focused laser heating and leads to a blue shift of the Raman frequency. When the stress effect is additionally considered, the measured frequency change $\Delta \omega$ would become

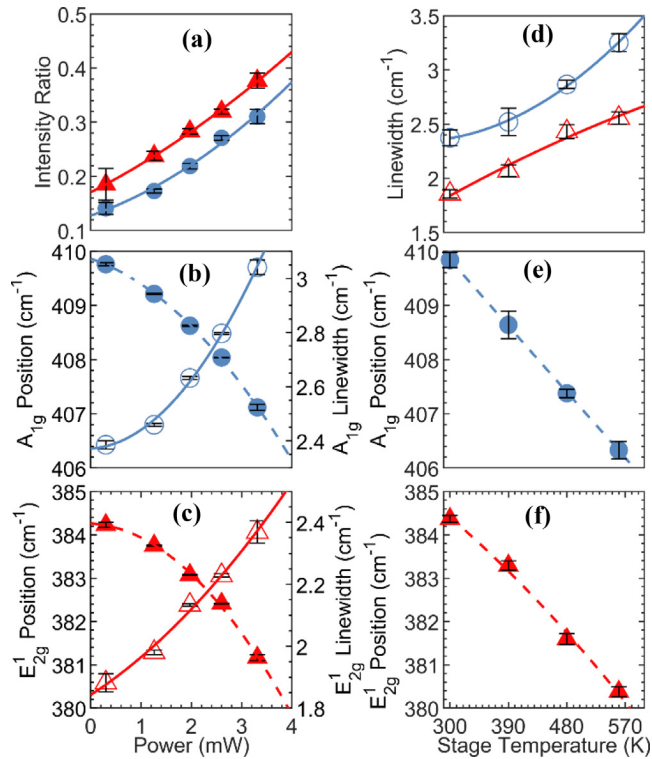

**FIG. 2.** (a) Anti-Stokes to Stokes intensity ratio and (b) and (c) the peak position (filled symbols) and linewidth (unfilled symbols) of the $A_{1g}$ (circles) and $E_{2g}^1$ (triangles) bands as a function of the incident laser power. (d) The extrapolated $A_{1g}$ (circles) and $E_{2g}^1$ (triangles) linewidth ($\Gamma_0$) and (e) and (f) peak position ($\omega_o$) at zero laser power as a function of the stage temperature. Each data point (symbols) represents the average of 10–15 spectra measured at a single power. The error bars are calculated based on the Student's T distribution at 95% confidence interval. Lines are second-order polynomial fits of the data. A 50× objective was used to obtain a Gaussian laser beam radius of 1.15 micron.

$$\Delta\omega(T,\sigma) = \left[\frac{\left(\frac{\partial \omega}{\partial \sigma}\right)_{\Delta T=0}}{\left(\frac{\partial \omega}{\partial T}\right)_{\sigma=0}} \beta(w_o) + 1\right] \left(\frac{\partial \omega}{\partial T}\right)_{\sigma=0} \Delta T, \quad (2)$$

where $\left(\frac{\partial \omega}{\partial \sigma}\right)_{\Delta T=0}$ is the linear coefficients of frequency change due to stress alone and is evaluated here from a reported Raman peak position of MoS$_2$ inside a diamond anvil cell.[33] $\beta(w_o)$ is the ratio between the lattice temperature rise and the stress. Due to nonuniform thermal stress inside the laser spot, the different stress-reduced peak shifts at different locations lead to an additional broadening of the measured Raman linewidth.[29]

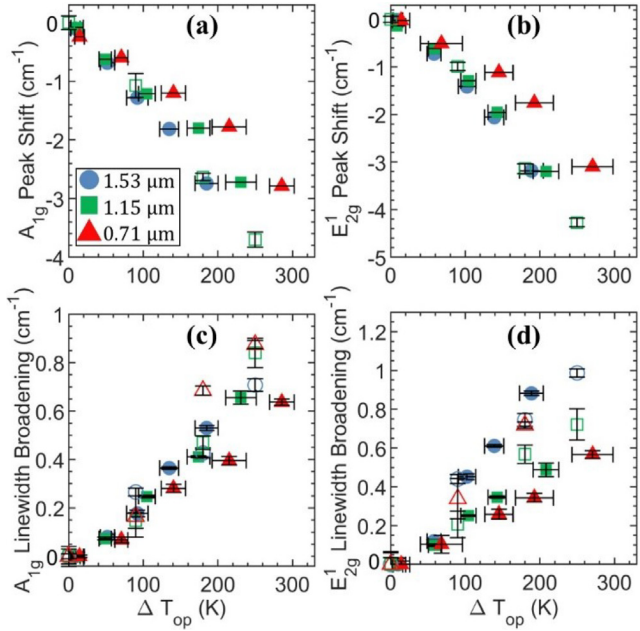

**FIG. 3.** (a) and (b) Measured frequency shift and (c) and (d) linewidth broadening vs the optical phonon temperature rise causes by the Raman laser heating for the (a) and (c) out-of-plane $A_{1g}$ and (b) and (d) in-plane $E_{2g}^1$ bands and different Gaussian beam width ($w_o$) values of 0.71 (triangles), 1.15 (squares), and 1.53 $\mu$m (circles). The filled symbols are data measured at different incident laser power and 300 K sample stage temperature. The unfilled symbols are obtained by extrapolating the measurement data at different laser power to zero laser power at each stage temperature.

By treating bulk $MoS_2$ as mechanically coupled monolayers, we calculate and plot the stress field caused by the focused laser heating, as shown in Fig. 4(a). Based on the Gaussian average of the lattice temperature rise and stress, we calculate $\beta(w_o)$ to be $6.4 \times 10^{-4}$, $6.5 \times 10^{-4}$, and $6.7 \times 10^{-4}$ GPa/K for $w_o$ of 1.53, 1.15, and 0.71 $\mu$m, respectively. Based on the calculated $\beta(w_o)$, we find that the stress-corrected peak shift temperature is about 17% and 8% higher than the peak shift and linewidth temperatures calculated without the stress correction for a beam waist of 0.71 $\mu$m, as shown in Fig. 4(b). These two values clearly and barely exceed the random uncertainty in the measurement.

At the smallest Gaussian beam radius of 0.71 $\mu$m, Fig. 4(b) shows that the stress-corrected peak shift temperature rise is about 32% lower than the optical phonon temperature rise determined from the intensity ratio. As the Gaussian beam radius is increased to 1.15 and 1.53 $\mu$m [Figs. 4(c) and S6], respectively, the difference between the two temperatures is reduced to 11% and comparable to the measurement uncertainty, indicating a thermalization length close to 1 $\mu$m between the hot optical phonons and the lattice. These results show less pronounced non-equilibrium between the zone-center optical phonons and the lattice than those observed previously in suspended monolayer graphene,[10] where the stress is readily relaxed. The observed moderate phonon non-equilibrium agrees with a first-principle based multi-temperature model.[22] That theoretical calculation shows that the largest nonequilibrium exists between the zone-center optical phonons and the other optical phonon modes in the Brillouin zone and the acoustic phonons, while there is only small nonequilibrium between optical and acoustic branches away from the zone center.

These results show that thermally induced compressive stress in multi-layer $MoS_2$ flake causes an observable blue shift of the phonon frequency obtained from micro-Raman measurements. This stress effect needs to be accounted for in the lattice temperature rise obtained from the Raman peak position and linewidth in Raman-based thermal transport measurements of multi-layer 2D samples, where the thermal stress cannot be readily relaxed. With this stress effect accounted for, the obtained peak shift temperature is 32% lower than the optical phonon temperature rise when the Raman laser spot size is reduced to 0.71 $\mu$m and below the 1 $\mu$m scale thermalization length. This latter finding agrees with a first principles theoretical prediction of a moderate energy dissipation bottleneck between the zone-center optical phonons and other phonon modes, instead of between the optical and acoustic phonon polarizations over the entire Brillouin zone. These insights are necessary for understanding both Raman-based thermal transport measurements and the energy dissipation pathways in 2D layered materials. This work serves as an initial step toward a comprehensive understanding of the interplay between highly nonequilibrium thermal transport and complicated thermomechanical phenomena, including intriguing thermally induced buckling instability in suspended monolayers, in 2D materials and devices.

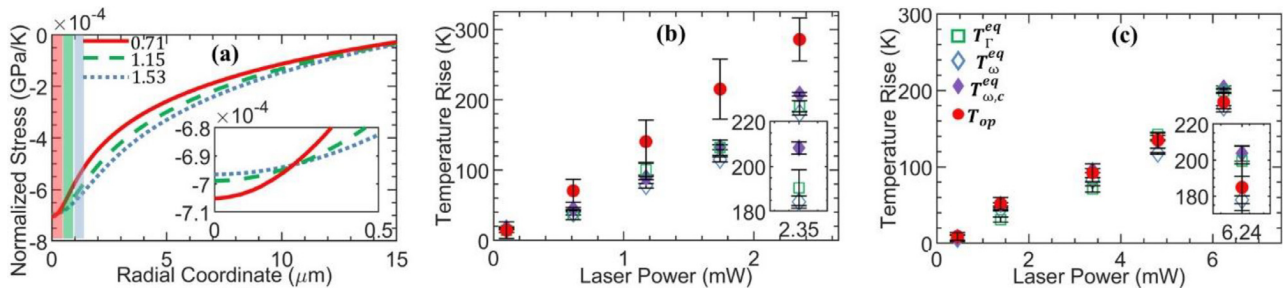

**FIG. 4.** (a) Calculated normalized compressive strain in a 72 nm-thick suspended $MoS_2$ flake as a function of the radial distance from the center of the laser spot with the Gaussian laser beam radius being 0.71 (red solid line), 1.15 (green dashed line), and 1.53 $\mu$m (blue dotted line). The profiles are normalized by the Gaussian weighted average temperature rise. (b) and (c) Temperature rise calculated from the $A_{1g}$ Raman intensity ratio ($T_{op}$, circles), linewidth ($T_{\Gamma}^{eq}$, squares), and peak shifts with ($T_{\omega,c}^{eq}$, filled diamonds) and without ($T_{\omega}^{eq}$, unfilled diamonds) stress correction as a function of laser power at the Gaussian beam width ($w_o$) values of 0.71 (b) and 1.53 $\mu$m (c). The stage temperature is 300 K. The legend in panel (c) applies to panel (b).

See the supplementary material for optical setup, Raman data, temperature measurement results at different laser spot sizes, and stress–strain calculations.

We thank Professor Chad Landis for helpful discussion of thermal stress calculation. L.S., P.S., X.R., and Z.H. were supported by two collaborating grants (Nos. CBET-2015946 and CBET-2015954) of the U.S. National Science Foundation (NSF). G.E. and B.S. conducted preliminary studies on samples provided by Michael Rodder with NSF support under Cooperative Agreement No. EEC-1160494.

## AUTHOR DECLARATIONS
### Conflict of Interest

The authors have no conflicts to disclose.

### Author Contributions

Peter Sokalski was guided by Li Shi to obtain and process all Raman data reported in the manuscript. Zherui Han was guided by Xiulin Ruan and Rui Huang to carry out multi-temperature and stress-strain calculations, respectively. Gabriella Coloyan Fleming was assisted by Sean Sullivan to obtain preliminary Raman data on the Raman setup built by Sean Sullivan. Brandon Smith obtained additional Raman data with a different setup. Li Shi and Xiulin Ruan conceived the concept and supervised the experimental and theoretical studies, respectively. All authors contributed to the writing and editing of the manuscript.

**Peter Sokalski:** Data curation (lead); Formal analysis (lead); Investigation (lead); Writing – original draft (lead); Writing – review & editing (equal). **Zherui Han:** Formal analysis (equal); Investigation (equal); Validation (equal); Writing – original draft (equal); Writing – review & editing (equal). **Gabriella Coloyan Fleming:** Methodology (equal). **Brandon Smith:** Data curation (equal); Investigation (equal). **Sean Sullivan:** Methodology (equal); Writing – review & editing (equal). **Rui Huang:** Formal analysis (equal); Investigation (equal); Supervision (equal); Writing – review & editing (equal). **Xiulin Ruan:** Conceptualization (lead); Funding acquisition (lead); Project administration (lead); Supervision (lead); Writing – review & editing (equal). **Li Shi:** Conceptualization (lead); Funding acquisition (lead); Investigation (equal); Project administration (lead); Supervision (lead); Writing – original draft (lead); Writing – review & editing (lead).

## DATA AVAILABILITY

The data that support the findings of this study are available from the corresponding author upon reasonable request.